\documentclass[reprint,prb,twocolumn,superscriptaddress,aps,longbibliography]{revtex4-2}

\usepackage[colorlinks=true,citecolor=blue]{hyperref} % hyperreferences
\usepackage{graphicx}
\usepackage{float} % Keep appendix figures with their discussion in each column.
\usepackage{bm}
\usepackage{amsmath}
\usepackage{amssymb}
\usepackage{comment}
\usepackage{braket}
\usepackage{orcidlink}
\usepackage{physics}
\UseRawInputEncoding

\DeclareGraphicsExtensions{.pdf,.png,.jpg,.eps}
\usepackage{xcolor}

\usepackage[normalem]{ulem}

\begin{document}

% \preprint{APS/123-QED}

\title{Tuning correlated states in rhombohedral pentalayer graphene with magnetic encapsulation}
%\title{\textbf{Programmable correlated electronic phases in magnetically proximitized rhombohedral pentalayer graphene}
%}%
%\title{\textbf{Ferromagnetic-pentalayer-chern manipulation?}

\author{Cuiju Yu\,\orcidlink{0000-0002-9726-7574}}
\thanks{These authors contributed equally}
\affiliation{
 Department of Applied Physics, Aalto University, 02150 Espoo, Finland
}

\author{Yitao Sun \,\orcidlink{0009-0002-9479-7147}}
\thanks{These authors contributed equally}
\affiliation{
 Department of Applied Physics, Aalto University, 02150 Espoo, Finland
}

\author{Jose L. Lado\,\orcidlink{0000-0002-9916-1589}}

\affiliation{
 Department of Applied Physics, Aalto University, 02150 Espoo, Finland
}

\date{\today}% It is always \today, today,

\begin{abstract}
Rhombohedral graphene multilayers provide a flexible platform
for correlated physics associated with their
flat-band electronic structure.
While a variety of correlated states appear in these systems as a function of bias and electrostatic
doping, promoting specific correlated flavor symmetry breaking states remains a challenge.
Here we show that magnetic encapsulation by CrGeTe$_3$ achieves this
in rhombohedral pentalayer graphene,
by using switchable exchange couplings induced by ferromagnetic substrates.
We show that correlations in rhombohedral multilayers at charge neutrality strongly renormalize the exchange proximity
splitting,
giving rise to collective spin modes whose character is switched by the parallel or antiparallel alignment of the encapsulation.
We further show that upon doping of the flat bands,
spin triplet $f$-wave superconducting solutions emerge,
whose layer and spin selectivity is controlled by the magnetic encapsulation,
hosting in-gap excitations at the edges.
Our results establish magnetic alignment as a means to control spin symmetry breaking and unconventional
superconductivity
in rhombohedral pentalayer graphene.
\end{abstract}

\maketitle

\section{Introduction}
Rhombohedral pentalayer graphene (R5G) has emerged as a highly tunable platform for correlated electron physics, owing to its nearly flat low-energy bands~\cite{Pierucci2015,zhang2026moire,https://doi.org/10.48550/arxiv.2511.17423}.
Experiments have revealed correlated insulating states~\cite{han2024correlated,PhysRevX.15.011045}, spin- and valley-polarized metallic phases~\cite{han2024correlated,Han2023}, integer and fractional Chern insulators~\cite{Han2024,Lu2024,75gl-jzl6,Lu2025} in R5G and R5G/hBN moir\'e systems.
More recently, unconventional superconductivity has been observed in R5G~\cite{han2025signatures,Seo2026}.
Similar to other van der Waals materials, the nature of the correlated states in R5G is not fully understood, due
to the competition of multiple correlated phases including charge density waves, spin ordered, valley ordered
and nematic phases~\cite{zhou2021half,Han2023,han2024correlated,75gl-jzl6,Dong2024,Soejima2024,Kwan2025}.
Hence, finding ways to favor specific correlated phases could help us understand how these phases interact.
% Hence, strategies to promote specific correlated phases would be highly desirable,
% which would potentially enable disentangling the interplay between the different correlated phases.
The flat bands in rhombohedral multilayers are strongly localized
in the outermost graphene layers, which makes them exceptionally tunable by proximity effects.
Proximity to ferromagnets can therefore tune the spin structure of low-energy bands and the resulting correlated states.

\begin{figure}[t!]
    \centering
    \includegraphics[width=\linewidth]{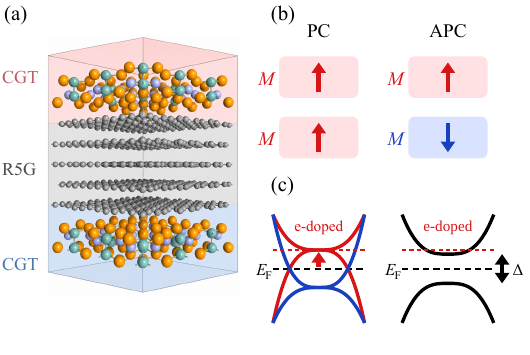}
    \caption{(a) Schematic of rhombohedral pentalayer graphene (R5G) encapsulated by two ferromagnetic CrGeTe$_3$ (CGT) monolayers.
    (b) The magnetizations $M$ of the two CGT layers are parallel in the PC (both pointing up) and antiparallel in the APC (top pointing up and bottom pointing down).
    (c) Schematic band structures of the PC, with band crossings near the Fermi level, and the APC, with a clear gap. A small gate voltage can shift the Fermi level upwards into the electron-doped regime.}
    \label{fig:Schematic}
\end{figure}

Proximity in van der Waals materials, including spin-orbit coupling
and exchange, represents a flexible strategy to create new emergent states.
Spin-orbit coupling proximity has been observed to promote specific correlated phases
in rhombohedral graphene~\cite{Zhang2023,patterson2025superconductivity,Yang2025,Han2024}
and twisted graphene multilayers~\cite{Bhowmik2022,Bhowmik2023}.
Magnetic exchange proximity effects have been demonstrated in graphene interfaced with several magnetic insulators. With yttrium iron garnet, the induced ferromagnetism is manifested as an anomalous Hall effect~\cite{PhysRevLett.114.016603}. With EuS, magnetotransport experiments reveal an interfacial exchange field exceeding 14~T~\cite{Wei2016}. With Cr$_2$Ge$_2$Te$_6$, the induced exchange gives rise to both an anomalous Hall effect and a spin-valve response~\cite{Yao2025,Yang2024}.
Beyond graphene, magnetic proximity has been shown to induce topological superconductivity in CrBr$_3$/NbSe$_2$ heterostructures~\cite{Kezilebieke2020} and a valley splitting of excitons in MoSe$_2$ stacked on CrBr$_3$~\cite{ciorciaro2020observation,Lyons2020}.
These strategies can be extended to full encapsulation
between two ferromagnetic layers, which provides additional control over the layer-resolved exchange
fields~\cite{PhysRevLett.121.067701,PhysRevB.104.075126,Soriano2020,Wolf2021}:
in bilayer graphene, the relative alignment of the two magnetizations yields a spin-valve effect, with a conducting state under parallel alignment and a gap opening near charge neutrality under antiparallel alignment~\cite{PhysRevLett.121.067701,PhysRevB.104.075126}.
Theoretical work on rhombohedral trilayer graphene further suggests
that proximity exchange can alter competing interaction-driven phases~\cite{PhysRevLett.132.186401}.
These developments motivate exploring how magnetic alignment affects the collective spin excitations of rhombohedral multilayers, as well as the spin selectivity and gap structure of their superconducting states.

Here, we show how magnetic alignment controls collective spin excitations and spin-selective superconducting pairing in R5G encapsulated between two ferromagnetic CrGeTe$_3$ layers.
Combining density functional theory (DFT) with an interacting tight-binding model, we show that parallel alignment of the magnetizations spin-splits the nearly flat surface bands, whereas antiparallel alignment induces a layer-staggered exchange potential that opens a gap at charge neutrality.
At half filling, electronic interactions enhance the proximity-induced exchange splitting, triggering
two different correlated phases.
These correlated states host collective spin modes that are absent in the non-interacting proximitized system.
Upon electron doping, residual attractive interactions
yield equal-spin triplet $f$-wave pairing whose spin selectivity depends on the magnetic configuration.
Parallel alignment supports pairing in only one spin channel, leaving the other gapless, and reversing the common magnetization switches the paired spin channel.
In contrast, antiparallel alignment enables pairing in both spin channels, producing a fully gapped quasiparticle spectrum.
We further compute armchair-edge spectra from these bulk Bogoliubov-de Gennes (BdG) Hamiltonians. 
The comparison between edge and bulk spectra reveals counterpropagating zero-energy-crossing subgap branches for parallel alignment, but nearly flat finite-energy subgap branches for antiparallel alignment.
Our results show that magnetic encapsulation offers a way to reconfigure collective correlated phases and spin selectivity of triplet superconducting orders in R5G.
The paper is organized as follows. In section~\ref{section:plain R5G}, we introduce the heterostructure and magnetic configurations,
using a combination of DFT and tight-binding descriptions of the low-energy bands.
In section~\ref{section:magnon}, we analyze the correlated magnetic phases driven by the proximity,
and the emergent collective modes arising from correlations at charge neutrality.
Section~\ref{section:SC} examines how magnetic alignment controls the spin selectivity and quasiparticle gaps of electron-doped equal-spin triplet $f$-wave superconducting states, and compares their armchair-edge and projected bulk spectra.
Finally, Section~\ref{sec:conclusion} summarizes the main conclusions.

% Defined on page 2 so that Fig. 2 floats to the top of page 3.
\begin{figure*}[tpb]
     \centering
    \includegraphics[width=\linewidth]{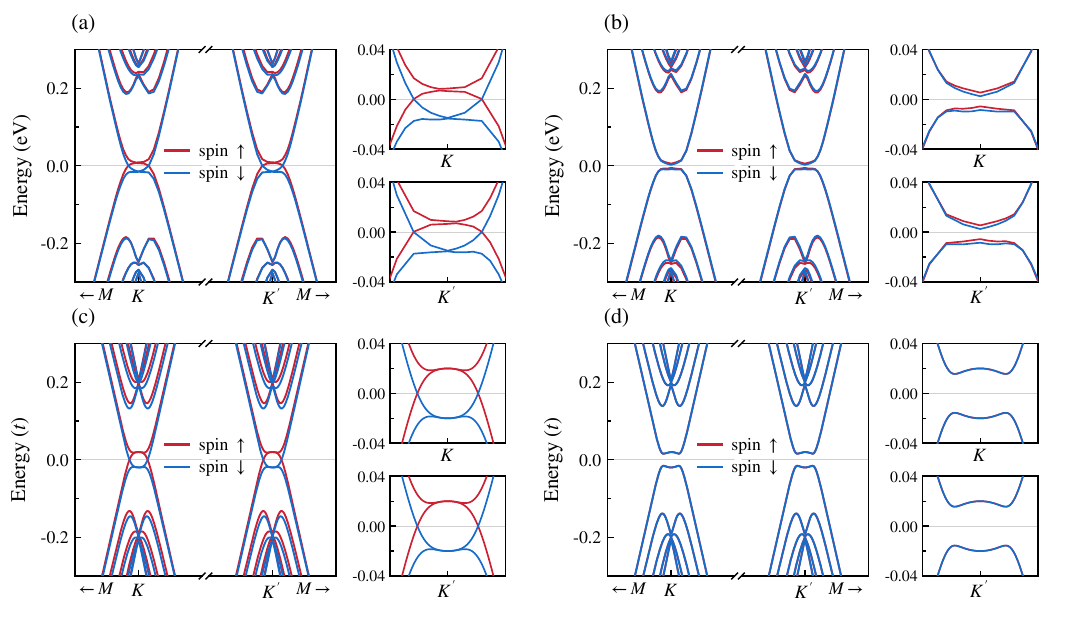}
    \caption{Spin-resolved electronic band structures of the CGT/R5G/CGT heterostructure, together with enlarged views around K and K$^\prime$ valleys, for the (a) PC and (b) APC obtained from DFT calculations. (c,d) Corresponding tight-binding spin-resolved band structures for (c) PC and (d) APC. The tight-binding bands in (c,d) are computed without interactions ($U=0$), with $J=0.02\,t$ and $t^{\perp}=0.2\,t$. In (d) the spin-up and spin-down bands are degenerate and overlap completely. Energies are in eV in (a,b) and in units of $t$ in (c,d).}
    \label{fig:bands}
\end{figure*}

\section{Magnetic proximity effects in rhombohedral pentalayer graphene}
\label{section:plain R5G}

\subsection{Pristine rhombohedral pentalayer graphene and magnetic configurations}

R5G consists of five graphene layers in ABCAB stacking (Fig.~\ref{fig:Schematic}a), whose low-energy states are mostly localized in the two outermost layers,
giving rise to a pair of spin- and valley-degenerate conduction and valence bands near charge neutrality~\cite{Guinea2006}.
Around the $K$ and $K^\prime$ valleys, these bands display a high-order chiral dispersion, leading to nearly flat bands at low energy~\cite{MinMacDonald2008},
whose bandwidth, overlap and gap can be modified by a perpendicular displacement field~\cite{Koshino2010} or by an exchange field.
As a result, R5G is highly sensitive to both magnetic proximity and interaction-driven instabilities.

In the following, we tune the electronic structure of R5G by encapsulating it between two layers of CGT,
a van der Waals ferromagnet that remains magnetic down to the few-layer limit~\cite{Gong2017,verzhbitskiy2020controlling}.
We perform DFT calculations for R5G with one CGT layer on top and one at the bottom, and for the optimized geometry of Fig.~\ref{fig:Schematic}a
we consider the two magnetic configurations of the CGT layers shown in Fig.~\ref{fig:Schematic}b.
In the layer-parallel configuration (PC) both magnetizations point up,
whereas in the layer-antiparallel configuration (APC) the top magnetization points up and the bottom one points down.
% Defined on page 3 so that Fig. 3 floats to page 4.
\begin{figure*}[t!]
    \centering
    \includegraphics[width=1.\linewidth]{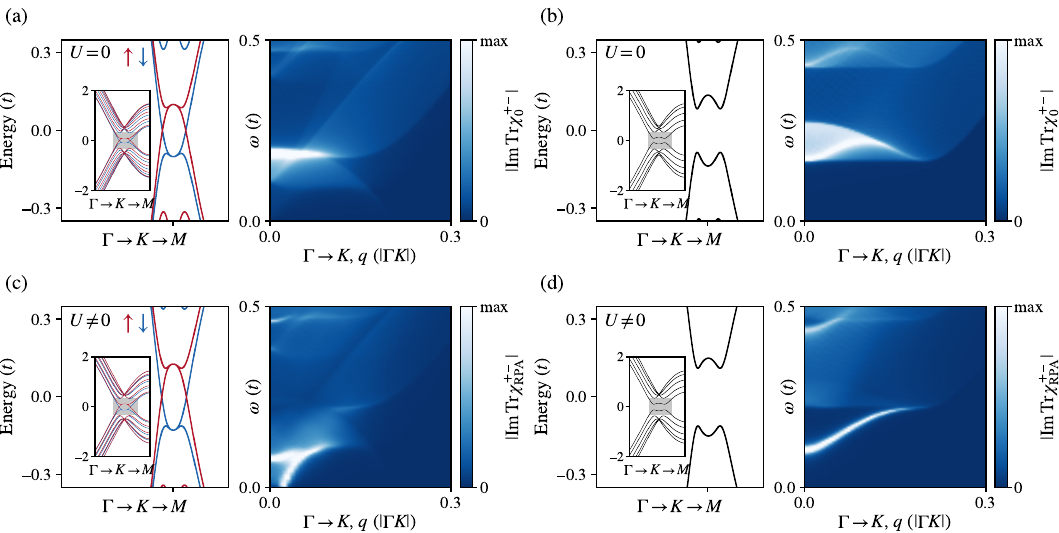}
    \caption{Band structures and transverse spin response of the PC (a,c) and APC (b,d) at half filling.
(a,b) Non-interacting cases ($U=0$), with exchange fields $J=0.1\,t$ (a) and $J=0.136\,t$ (b).
(c,d) Similar bands obtained with a weaker proximity exchange $J=0.05\,t$, with magnetism enhanced self-consistently by the Hubbard interaction $U=2.1\,t$.
Insets show the bands along $\Gamma\to K\to M$, starting at $0.45\,|\Gamma K|$, over a wider energy range; gray marks the main-panel window $|E|<0.35\,t$.
PC bands are red (spin up) and blue (spin down), while spin-degenerate APC bands are black.
The right side of each panel shows $|\mathrm{Im}\,\mathrm{Tr}\,\boldsymbol{\chi}^{+-}(q,\omega)|$, with the trace over sites, versus $q$ along $\Gamma\to K$ (in units of $|\Gamma K|$) and $\omega$ (in units of $t$).
Each map is normalized to its own 99th percentile, with higher values saturated.
Collective spin excitations appear only in the interacting cases (c,d); the non-interacting cases (a,b) show only the particle-hole continuum.}
    \label{fig:rpa_fe_af}
\end{figure*}
As sketched in Fig.~\ref{fig:Schematic}c, parallel alignment gives rise to spin-split bands crossing near the Fermi level,
whereas antiparallel alignment opens a gap at charge neutrality.
Upon electron doping, the Fermi level of the PC enters a spin-polarized region of the flat bands, whose spin polarization is reversed by reversing both magnetizations.
These two electronic structures are the starting point for the correlated phases of Sec.~\ref{section:magnon}
and for the triplet superconductivity of Sec.~\ref{section:SC}.

\subsection{Tight-binding model and band structures}
\label{subsec:TB}

To describe the low-energy bands of the heterostructure, we use a tight-binding model for the carbon $p_z$ orbitals of R5G~\cite{Pyqula},
in which the magnetic substrates enter only through a proximity exchange field.
The Hamiltonian is
\[
H  = H_0 + H_J  ,
\]
where
\begin{equation}
H_0 =
- \sum_{\langle i,j \rangle_{\parallel}, \sigma}
t^{\parallel}_{ij}\, c^\dagger_{i\sigma} c_{j\sigma}
- \sum_{\langle i,j \rangle_{\perp}, \sigma}
t^{\perp}_{ij}\, c^\dagger_{i\sigma} c_{j\sigma}
 .
\label{eq:H0}
\end{equation}
Here $t^{\parallel}_{ij}$ is the hopping between nearest neighbors in the same layer $\langle i,j\rangle_{\parallel}$,
$t^{\perp}_{ij}$ is the hopping between vertically adjacent sites of neighboring layers $\langle i,j\rangle_{\perp}$, and $\sigma = \{\uparrow,\downarrow \}$ is the spin.
We take $t^{\parallel}_{ij}=t$ and $t^{\perp}_{ij}=t^{\perp}$.
We neglect the remote interlayer hoppings, which give rise to trigonal warping and modify the bands at the lowest energies~\cite{KoshinoMcCann2009,Zhang2010},
and we also neglect spin-orbit coupling, so that we isolate the effect of the proximity exchange on the correlated and superconducting states.
The exchange term $H_{J}$ acts only on the top and bottom layers of R5G, which are in direct contact with the magnetic substrates,
\begin{equation}
H_{J}
= -J \sum_{\substack{i \in \ell_\mathrm{t},\, \ell_\mathrm{b} \\ \sigma\sigma'}}
\mathbf{m}_i \cdot
\left(
c^\dagger_{i\sigma}\, \boldsymbol{\sigma}_{\sigma\sigma'}\, c_{i\sigma'}
\right),
\label{eq:HJ}
\end{equation}
where $\ell_\mathrm{t}$ and $\ell_\mathrm{b}$ are the sites of the top and bottom layers, $\mathbf{m}_i$ is the direction of the induced exchange,
$J$ is the bare proximity exchange, and $\boldsymbol{\sigma}$ are the Pauli matrices.
We give all energies in units of the intralayer nearest-neighbor hopping, $t=1$, with $t=2.7$~eV.
We take the exchange fields collinear and perpendicular to the layers, $\mathbf{m}_i=m_\mathrm{t}\hat{\mathbf{z}}$ for $i\in\ell_\mathrm{t}$ and $\mathbf{m}_i=m_\mathrm{b}\hat{\mathbf{z}}$ for $i\in\ell_\mathrm{b}$,
with $m_\mathrm{t},m_\mathrm{b}=\pm1$, where $m_\mathrm{t}=m_\mathrm{b}$ in the PC and $m_\mathrm{t}=-m_\mathrm{b}$ in the APC. Spin-up and spin-down are defined relative to a fixed $z$ axis.
For $J>0$, $H_J$ lowers by $J$ the energy of an electron on an outer layer with spin parallel to $\mathbf{m}_i$ and raises by $J$ the energy of the antiparallel spin, leading to a bare onsite spin splitting of $2J$.

The DFT bands of the CGT/R5G/CGT heterostructure in Figs.~\ref{fig:bands}a,b show a clear difference between the PC and the APC:
in the PC, bands of opposite spin cross near both the $K$ and $K^\prime$ valleys, whereas in the APC a gap opens near charge neutrality
(the DFT details are given in Appendix~\ref{app:methods}).
The minimal tight-binding model reproduces this difference qualitatively, as shown in Figs.~\ref{fig:bands}c,d.
This difference stems from the effect of $H_J$, Eq.~\eqref{eq:HJ}, on the low-energy states, which are localized mostly in the outer layers.
In the PC the exchange has the same sign on both outer layers, and therefore acts in this subspace as a uniform Zeeman field,
shifting the spin-up and spin-down bands in opposite directions and leading to band crossings.
In the APC the exchange has opposite signs on the two outer layers, and this layer-staggered exchange acts as a spin-dependent mass term that opens a gap,
as in the van der Waals spin valve of encapsulated bilayer graphene~\cite{PhysRevLett.121.067701}.

\section{Correlated state and collective modes at charge neutrality}
\label{section:magnon}

Electronic interactions in the nearly flat low-energy bands amplify the proximity-induced exchange splitting and stabilize correlated phases that depend on the magnetic configuration.
Among the many correlated states that compete in R5G, we focus on the two that the magnetic encapsulation promotes,
namely a magnetic state at charge neutrality and a spin-triplet superconducting state upon doping, which we address in Sec.~\ref{section:SC}.
We focus on these two because the proximity exchange directly breaks the spin degeneracy of the flat bands, thus selecting the spin channel,
whereas charge, valley and nematic orders remain outside our collinear mean-field description.
We now include
electronic interactions in the model by means of a minimal
local Hubbard interaction as $H=H_0+H_J+H_U$, with $H_U=U\sum_i n_{i\uparrow}n_{i\downarrow}$,
where \(n_{i\sigma}=c^\dagger_{i\sigma}c_{i\sigma}\) and $i$ labels the sites.
Assuming a collinear magnetic state at site $i$, we decouple the Hubbard interaction at the mean-field level as

\begin{equation}
H_U^{\rm MF} = U\sum_i \left( \langle n_{i\uparrow}\rangle n_{i\downarrow} + \langle n_{i\downarrow}\rangle n_{i\uparrow} - \langle n_{i\uparrow}\rangle \langle n_{i\downarrow}\rangle \right),
\label{eq:HU_MF}
\end{equation}
introducing the local charge density $n_i=\langle n_{i\uparrow}\rangle+\langle n_{i\downarrow}\rangle$ and spin polarization $m^z_i=\langle n_{i\uparrow}\rangle-\langle n_{i\downarrow}\rangle$.
Up to a constant, the mean-field interaction can be rewritten as
\begin{equation}
 H_U^{\rm MF} = \sum_{i,\sigma,\sigma'} \left[ \mu_i^U \delta_{\sigma\sigma'} - J_i^U (\sigma_z)_{\sigma\sigma'} \right] c^\dagger_{i\sigma}c_{i\sigma'},
\end{equation}
with
$ \mu_i^U=\frac{Un_i}{2}$, $J_i^U=\frac{Um^z_i}{2}$.
The first term represents an interaction-induced shift of the local chemical potential, whereas the second acts as an additional exchange field.
When the interaction-induced polarization is aligned with the
proximity field, the Hubbard interaction therefore enhances the effective spin splitting,
as shown in Figs.~\ref{fig:rpa_fe_af}c,d, where a small proximity exchange amplified by the interaction
gives rise to a splitting comparable to the one obtained with a two to three times larger exchange in the absence of interactions (Figs.~\ref{fig:rpa_fe_af}a,b).
As a result, the PC realizes a ferromagnetic metal, and the APC a layer-antiferromagnetic insulator with opposite spin polarization on the two surfaces.
While interactions renormalize the value of the exchange splitting,
a splitting driven by proximity and one driven by interactions cannot be distinguished from the
band structure alone.
Collective excitations, as measured by the dynamical spin structure factor,
in contrast directly reflect the different origin of the spin splitting.
With interactions, the spin response can also contain collective excitations distinct from single-particle excitations. These collective excitations may form coherent modes below the particle–hole continuum.
In the non-interacting case, the spin response instead consists of particle-hole excitations determined by the underlying single-particle electronic structure~\cite{henriques2024beyond,catarina2023broken,PhysRevLett.124.046403,PhysRevB.111.075103}.
To isolate this distinction, we compare two scenarios with similar low-energy band structures: a non-interacting system subject to a larger imposed exchange field, and an interacting system in which a smaller proximity exchange is amplified by the Hubbard interaction at the mean-field level.

The interaction-induced collective spin dynamics can be characterized by
the transverse spin susceptibility within the random phase approximation (RPA)~\cite{VeraMontoto2026,Yu2026altermagnet}.
Starting from the self-consistent mean-field state, the RPA
susceptibility is obtained by

\begin{equation}
\boldsymbol{\chi}^{+-}_{\rm RPA}(\mathbf q,\omega)
=
\left[
\mathbf I+U\boldsymbol{\chi}^{+-}_0(\mathbf q,\omega)
\right]^{-1}
\boldsymbol{\chi}^{+-}_0(\mathbf q,\omega),
\end{equation}

where $\boldsymbol{\chi}^{+-}_0$ is the bare transverse spin susceptibility
evaluated using the mean-field eigenstates,

\begin{equation}
\chi^{+-}_{0,\alpha\beta}(\mathbf q,\omega)
=
\frac{1}{N_k}
\sum_{\mathbf k\mu\nu}
\frac{f_{\mu\mathbf k}-f_{\nu\mathbf k'}}
{\epsilon_{\mu\mathbf k}-\epsilon_{\nu\mathbf k'}-\omega+i\eta}
M^{\alpha}_{\mu\nu}M^{\beta*}_{\mu\nu}.
\end{equation}

Here, $\mu$ and $\nu$ label the mean-field bands, with energies $\epsilon_{\mu\mathbf k}$ and eigenstates $|\psi_{\mu\mathbf k}\rangle$,
$f_{\mu\mathbf k}=f(\epsilon_{\mu\mathbf k})$ is the Fermi--Dirac occupation,
$\alpha$ and $\beta$ denote the site degrees of freedom, and
$M^{\alpha}_{\mu\nu}=\langle\psi_{\mu\mathbf k}|S^+_{\alpha}|\psi_{\nu\mathbf k'}\rangle$
is the spin-flip matrix element, with $\mathbf k'=\mathbf k+\mathbf q$ and $S^\pm_{\alpha}$ the local spin
raising and lowering operators.
We use a susceptibility broadening $\eta=3\times10^{-3}\,t$, except for the non-interacting APC in Fig.~\ref{fig:rpa_fe_af}b, where $\eta=2\times10^{-3}\,t$.
The Fermi--Dirac occupation smearing energy in the susceptibility calculation is set equal to $\eta$.
Collective spin excitations correspond to poles of the RPA susceptibility, and in the vanishing-broadening limit an undamped mode satisfies $\det[\mathbf I+U\boldsymbol{\chi}^{+-}_0(\mathbf q,\omega)]=0$.

For $U=0$ the RPA susceptibility reduces to the bare bubble $\boldsymbol{\chi}^{+-}_0$, which describes spin-flip particle-hole excitations without collective modes,
and therefore the spectra in Figs.~\ref{fig:rpa_fe_af}a,b are the non-interacting references for those in Figs.~\ref{fig:rpa_fe_af}c,d.
While the single-particle bands are similar in the two cases, the transverse spin responses are qualitatively different.
In the PC, the gapless band crossings give rise to a broad Stoner continuum extending down to low frequencies,
and the Hubbard interaction leads to the emergence of dispersive collective spin resonances,
featuring a soft low-energy branch that partially overlaps with the continuum (Fig.~\ref{fig:rpa_fe_af}c).
This overlap provides decay channels for the collective excitations, which are therefore subject to Landau damping~\cite{Buczek2011}.
In contrast, the APC has a finite threshold for spin-flip particle-hole excitations,
and its interacting response displays a gapped collective branch below the Stoner continuum that remains sharp over a broad momentum range (Fig.~\ref{fig:rpa_fe_af}d).
Since this branch is separated from the continuum, Landau damping is suppressed and the mode remains well defined. We refer to this branch as the magnon mode of the system.
At fixed $t^\perp$, $U$ and $J$, switching between parallel and antiparallel magnetizations therefore changes both the low-energy bands and the collective spin excitations, together with their damping.
Inelastic electron tunneling through vertical van der Waals junctions, which has resolved the magnons of CrI$_3$ and CrBr$_3$~\cite{Klein2018,Ghazaryan2018}, could probe these modes and their switching.

\section{Spin-triplet superconductivity controlled by the exchange fields}
\label{section:SC}

\begin{figure}[!t]
    \centering
    \includegraphics[width=\columnwidth]{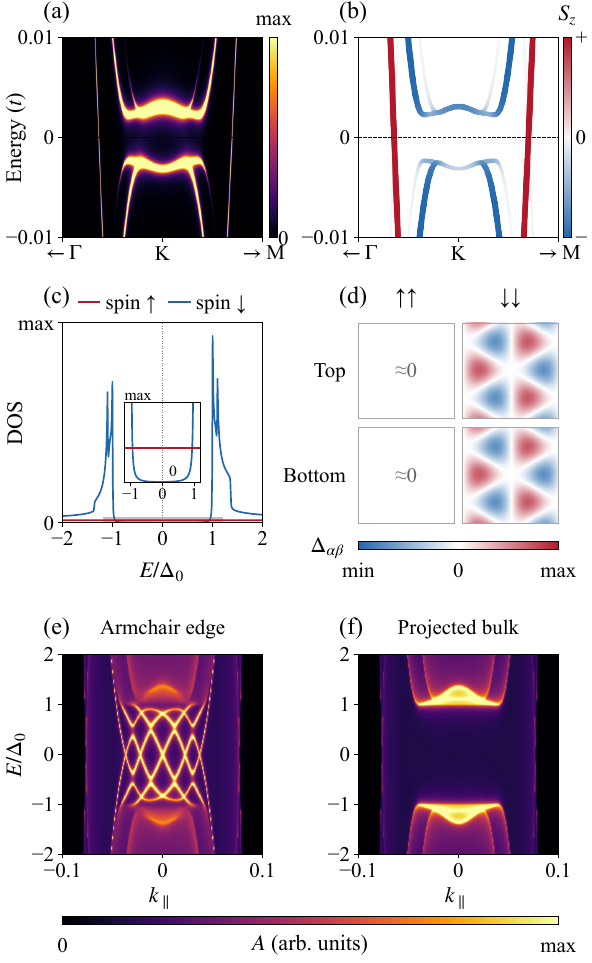}
    \caption{Superconducting state in the parallel configuration (PC).
(a) Momentum-resolved spectral function.
(b) Band structure colored by $S_z$.
(c) Spin-resolved DOS, with the gray region enlarged in the inset.
(d) Equal-spin pairing amplitudes $\operatorname{Re}\Delta_{\alpha\alpha}(\mathbf{k})$ for the top and bottom layers (rows) and the $\uparrow\uparrow$ and $\downarrow\downarrow$ channels (columns). The gap is diagonal in the orbital index. Only spin-down electrons pair, opening a gap in the spin-down sector while the spin-up sector remains gapless. After fixing a single global phase for each magnetic configuration, all plotted pairing components are real to numerical precision.
(e,f) Spectral weight $A$ [Eq.~(\ref{eq:spectral})] at an armchair edge (e) and in the bulk projected onto the same momentum $k_\parallel$ (f). Both panels use a square-root color scale. Subgap branches crossing zero energy appear only in the edge spectrum.
Energies in (c,e,f) are given in units of $\Delta_0=2.3\times10^{-3}\,t$, the minimum positive BdG energy in the spin-down sector.}
    \label{fig:superconducting}
\end{figure}

We now turn to superconductivity under electron doping, keeping the Hubbard term $H_U$ and adding a residual attraction $V_2<0$ between next-nearest-neighbor (NNN) sites.
We take this channel as the dominant one because the low-energy surface states are almost sublattice polarized, and NNN sites belong to the same sublattice~\cite{VinasBostrom2024}.
We take $V_2$ as a phenomenological attraction, and we do not address its microscopic origin,
whether phonons, spin fluctuations or the anyon-mediated pairing proposed for R5G near its fractional Chern states~\cite{Shi2025,Kim2025,Divic2025,wang2025chiral,Guerci2025}.
The Hamiltonian is
\begin{equation}
H = H_0 + H_J + H_U + V_2 \sum_{\substack{\ell,\langle\!\langle i,j\rangle\!\rangle \\ \sigma,\sigma'}} n_{i\ell\sigma} n_{j\ell\sigma'} ,
\end{equation}
where $\ell$ is the layer index, $\langle\!\langle i,j\rangle\!\rangle$ are the NNN pairs in each layer, and $n_{i\ell\sigma}=c^\dagger_{i\ell\sigma}c_{i\ell\sigma}$.
We decouple both interactions in all the Wick channels, namely charge, spin, hopping renormalization and pairing,
and solve the resulting BdG Hamiltonian self-consistently, with the pairing channel giving the bond order parameter $\Delta_{ij,\ell}^{\sigma\sigma'} = V_2 \langle c_{j\ell\sigma'}c_{i\ell\sigma}\rangle$.

Each site belongs to a unit cell $\mathbf{R}_i$ and to an orbital $\alpha=(s,\ell)$, where $s$ is the sublattice.
In momentum space the gap matrix is
\begin{equation}
\left[\hat{\Delta}_{\alpha\beta}(\mathbf{k})\right]_{\sigma\sigma'} = \sum_{j\in\beta} \Delta_{ij,\ell}^{\sigma\sigma'}\, e^{i\mathbf{k}\cdot(\mathbf{R}_j-\mathbf{R}_i)},
\end{equation}
where $i$ is the site of orbital $\alpha$ in the unit cell at the origin, and the sum runs over the sites $j$ of orbital $\beta$ in all the unit cells.
This is a $2\times2$ matrix in spin for each pair of orbitals, and since NNN bonds join sites of the same sublattice and the same layer,
$\hat{\Delta}_{\alpha\beta}(\mathbf{k})$ vanishes for $\alpha\neq\beta$.
We write each orbital component in terms of a spin-singlet and a spin-triplet part,
\begin{equation}
\hat{\Delta}_{\alpha\alpha}(\mathbf{k})
=
\left[
\psi_{\alpha}(\mathbf{k})\sigma_0
+
\mathbf{d}_{\alpha}(\mathbf{k})\cdot\boldsymbol{\sigma}
\right] i\sigma_y .
\label{eq:gap_decomposition}
\end{equation}
Fermionic antisymmetry, $\hat{\Delta}_{\alpha\alpha}(\mathbf{k})=-\hat{\Delta}^{T}_{\alpha\alpha}(-\mathbf{k})$, makes $\psi_\alpha$ even and $\mathbf{d}_\alpha$ odd in $\mathbf{k}$.
The in-plane components of $\mathbf{d}_\alpha$ give the equal-spin pairing, $\Delta^{\uparrow\uparrow}_{\alpha}=-d_{\alpha,x}+id_{\alpha,y}$ and $\Delta^{\downarrow\downarrow}_{\alpha}=d_{\alpha,x}+id_{\alpha,y}$.
The singlet $\psi_\alpha$ and the component $d_{\alpha,z}$ give the opposite-spin pairing.

We use $U=2.0\,t$, $J=0.02\,t$, $V_2=-0.3\,t$ and $k_BT=10^{-4}\,t$ for all the magnetic configurations,
and we compare the configurations at the same converged density, $1.1\times10^{-2}$ extra electrons per primitive cell,
for which the Fermi level of the PC lies in the spin-polarized flat band.
In all the converged states the pairing is mostly equal-spin pairing, as expected from the spin polarization of the flat surface bands.

\begin{figure}[!t]
    \centering
    \includegraphics[width=\columnwidth]{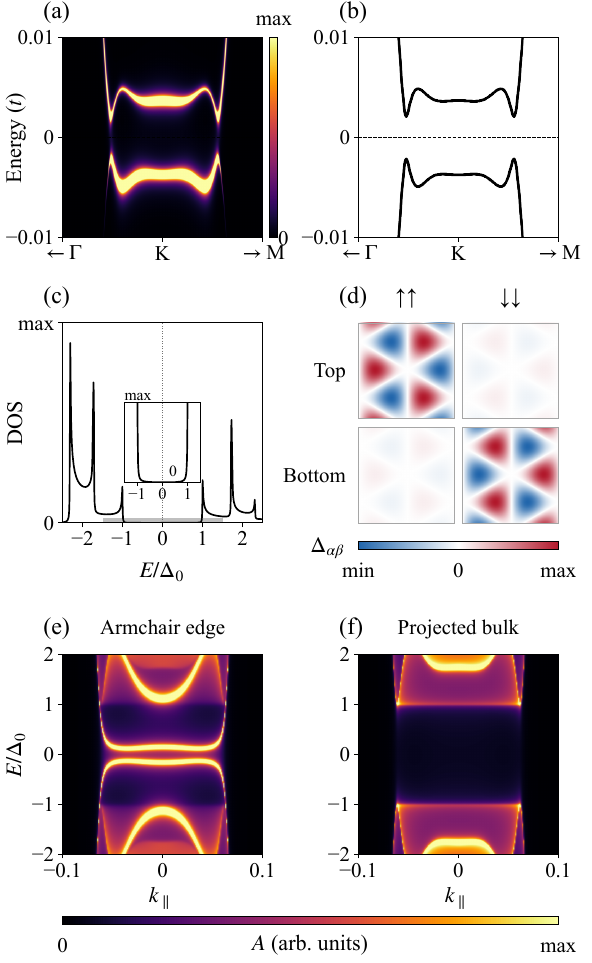}
    \caption{Superconducting state of the APC, with the panels of Fig.~\ref{fig:superconducting}.
    (a) Momentum-resolved spectral function and (b) spin-degenerate band structure.
    (c) Total DOS. Energies in (c,e,f) are in units of $\Delta_0=2.1\times10^{-3}\,t$, the minimum positive BdG energy, the same in both spin sectors.
    (d) Equal-spin pairing on the color scale of Fig.~\ref{fig:superconducting}d: spin up pairs mainly on the top layer and spin down mainly on the bottom layer.
    Both spin channels pair, so the quasiparticle spectrum is fully gapped.
    (e,f) Spectral weight $A$, Eq.~(\ref{eq:spectral}), at an armchair edge (e) and in the projected bulk (f), on the scale of Figs.~\ref{fig:superconducting}e,f.
    The edge hosts nearly flat subgap branches at finite energy.}
    \label{fig:SC}
\end{figure}

The edge spectrum directly reflects the sign structure of the gap.
Since the $f$-wave gap is odd in $\mathbf{k}$ and $K^\prime=-K$, it has opposite sign in the two valleys.
An armchair edge projects $K$ and $K^\prime$ onto the same $k_\parallel$, so edge scattering mixes the two valleys, and the sign change of the gap generates subgap states.
We identify edge branches by comparing the edge and projected bulk spectra at the same $k_\parallel$. In the PC, these branches lie within the gap of the paired spin-down sector, alongside a gapless spin-up bulk continuum while in the APC, they lie within the full bulk gap.
The spectral weight in the edge and projected bulk panels of Figs.~\ref{fig:superconducting} and \ref{fig:SC} is
\begin{equation}
A(k_\parallel,\omega)=\sum_{\alpha}\langle\alpha|\,\delta(\omega-H_{k_\parallel})\,|\alpha\rangle,
\label{eq:spectral}
\end{equation}
where $H_{k_\parallel}$ is the BdG Hamiltonian at fixed momentum $k_\parallel$ along the edge,
and the sum runs over the electron and hole orbitals of both spins on all five graphene layers,
taken in the outermost principal layer for the edge and in one bulk unit cell for the projected bulk.
The edge is obtained from the surface Green's function of a semi-infinite system with an armchair edge,
keeping the normal and pairing mean fields at their bulk values up to the boundary, and therefore the edge is not solved self-consistently.

We first consider the superconducting PC with $m_\mathrm{t}=m_\mathrm{b}=+1$, opposite to the orientation $m_\mathrm{t}=m_\mathrm{b}=-1$ used for the normal-state PC spectra in Figs.~\ref{fig:bands}c and \ref{fig:rpa_fe_af}a,c. This reversal interchanges the roles of spin-up and spin-down.
At this density only the spin-down flat band lies at the Fermi level, leading to a gap opening solely in the spin-down channel (Figs.~\ref{fig:superconducting}a--c),
whereas the spin-up Fermi surface stems from strongly dispersing bands with a much smaller density of states and remains gapless.
Therefore, the total spectrum is gapless, and the pairing is spin selective because the flat bands are spin selective.
The spin-down order parameter has the same amplitude on the two outer layers (Fig.~\ref{fig:superconducting}d),
and it displays equal-spin triplet $f$-wave symmetry, odd in $\mathbf{k}$, with nodal lines along $\Gamma$--$M$ far from the Fermi pockets at $K$ and $K^\prime$,
which is the pairing expected for an NNN attraction in the honeycomb lattice~\cite{Honerkamp2008,Chou2021,SzaboRoy2022}.
The projected bulk spectrum shows the spin-down gap and a weak continuum stemming from the gapless spin-up bands (Fig.~\ref{fig:superconducting}f),
whereas at the armchair edge spin-down subgap branches cross zero energy (Fig.~\ref{fig:superconducting}e).
These branches originate from edge scattering between $K$ and $K^\prime$, where the $f$-wave gap has opposite sign. They propagate in opposite directions and are therefore not chiral.
Reversing both exchange fields interchanges the roles of the two spins, so that the other spin channel pairs, as we verified through a self-consistent bulk calculation. Overall, the PC acts as a superconducting spin filter controlled by the magnetization.
% Queue the APC figure later to separate it from the PC figure.

In the APC we take $m_\mathrm{t}=-1$ and $m_\mathrm{b}=+1$, so the exchange field has opposite sign on the two outer layers, and the two spins remain degenerate.
The normal-state conduction flat bands shown in Figs.~\ref{fig:bands}b,d have opposite-spin components localized mainly on opposite outer layers. At the chosen electron doping, both spin channels cross the Fermi level and pair, yielding a fully gapped quasiparticle spectrum, as shown in Figs.~\ref{fig:SC}a--c, in contrast to the gapless PC spectrum in Figs.~\ref{fig:superconducting}a--c.
The pairing remains equal-spin $f$-wave, with spin up dominating on the top layer and spin down on the bottom layer (Fig.~\ref{fig:SC}d),
the other spin component being much weaker on each layer.
The armchair edge hosts nearly flat subgap branches with equal spectral weight in both spin channels (Figs.~\ref{fig:SC}e,f). As in the PC, these branches arise from the sign change of the $f$-wave gap between the two valleys mixed by the edge.
Their lowest peaks, however, remain at finite energy, well inside the projected bulk gap.
The two edge spectra therefore show different features, with zero-energy crossings in the PC and finite-energy branches in the APC, providing a signature of the underlying magnetic order.
Spin-polarized tunneling~\cite{Tedrow1982,Cornils2017,vanDriel2023} could probe the spin dependence of the gap and its switching with magnetization in the PC, while spatially resolved tunneling at the armchair edge~\cite{Menard2017,PalacioMorales2019,Kezilebieke2020} could distinguish the subgap spectra of the two configurations.

\section{Conclusion}
\label{sec:conclusion}

In summary, we have shown that the relative alignment of the magnetizations of two encapsulating CrGeTe$_3$ layers can favor specific correlated states in rhombohedral pentalayer graphene, a system with many competing orders.
DFT and tight-binding calculations show that the proximity exchange spin-splits the nearly flat surface bands in PC and opens a gap in APC.
At charge neutrality the Hubbard interaction amplifies this splitting, giving a ferromagnetic metal and a layer-antiferromagnetic insulator, respectively.
These correlated states host collective spin modes which are absent for a single-particle exchange field producing the same splitting.
In the PC, the modes overlap with the Stoner continuum, but in the antiparallel configuration, a sharp gapped magnon emerges below it.
Upon electron doping, a residual next-nearest-neighbor attraction yields equal-spin triplet $f$-wave superconductivity whose spin and layer structure is set by the magnetic configuration.
PC pairs only one spin channel, selected by the direction of the common magnetization,
whereas APC pairs both spin channels, with opposite dominant spins on the two surfaces, leading to a fully gapped quasiparticle spectrum.
The armchair edge hosts zero-energy crossings in the first case and finite-energy subgap branches in the second.
% The collective modes could be probed by inelastic tunnelling with van der Waals electrodes, and the spin-resolved gaps and edge spectra by tunnelling spectroscopy.
Magnetically encapsulated rhombohedral multilayers thus allow both collective spin modes and spin-selective triplet superconductivity to be switched by the magnetization.

\begin{acknowledgments}
We acknowledge the computational resources provided by the Aalto Science-IT project and the financial support from  InstituteQ,
the Jane and Aatos Erkko Foundation, the Academy of Finland Projects No. 370912, the Finnish Ministry
of Education and Culture through the Quantum Doctoral Education Pilot Program (QDOC VN/3137/2024-OKM-4), the Research Council of Finland through the
Finnish Quantum Flagship project (358877, Aalto University),
the Finnish Centre of Excellence in Quantum Materials QMAT (No. 374166),
and the ERC Consolidator Grant ULTRATWISTROICS (Grant agreement no. 101170477). We thank A. Moustaj, T. Ant\~ao, and A. Fumega for useful discussions.
\end{acknowledgments}

\appendix

\raggedbottom

\section{Computational details}
\label{app:methods}

We carry out the density functional theory (DFT) calculations with the Perdew-Burke-Ernzerhof (PBE) generalized gradient approximation (GGA) as implemented in the Quantum ESPRESSO (QE) package~\cite{Perdew1996,Giannozzi2009}, with optimized norm-conserving Vanderbilt (ONCV) pseudopotentials~\cite{hamann2013optimized,hamann2017erratum} and the DFT-D3 dispersion correction of Grimme~\cite{grimmeConsistentAccurateInitio2010} for the van der Waals interactions. We use an energy cutoff of 500 eV and a $9\times9\times1$ $k$-point mesh for the supercell~\cite{monkhorst1976special}, and a vacuum region of about 11~{\AA} along the $c$ axis ($c = 40$~{\AA}) to decouple the periodic images of the slab. We model the CrGeTe$_3$ (CGT)/R5G/CGT heterostructure with a $\sqrt{7}\times\sqrt{7}$ graphene supercell on a $1\times1$ CGT cell (90 atoms) at an in-plane lattice constant of 6.87~{\AA}, close to our PBE-D3 value for CGT (6.86~{\AA}) and to the experimental value (6.83~{\AA})~\cite{carteaux1995crystallographic}. The lattice mismatch with PBE graphene (2.47~{\AA}~\cite{buda2017characterization}) is about 5\%, so we stretch the graphene layers uniformly to match CGT. This biaxial strain preserves the $C_3$ symmetry and the $K$/$K^\prime$ valleys and does not enter the tight-binding model. We relax the atomic positions at fixed lattice parameters. We apply no Hubbard correction to the Cr $d$ orbitals, since the CGT magnetic moments are already stable at the plain PBE level.

% Use the APS column-width reference divider.
\makeatletter
\renewcommand{\bibsection}{%
  \par
  \begingroup
    \baselineskip26\p@
    \bib@device{\hsize}{72\p@}%
  \endgroup
  \nobreak\@nobreaktrue
  \addvspace{19\p@}%
}
\makeatother

\bibliography{biblio}

\end{document}